\documentclass{ifacconf}
\usepackage{graphicx}      
\usepackage{natbib}        
\usepackage{amsmath,amssymb,amsfonts,bm,mathtools}
\usepackage{comment}
\usepackage{siunitx}
\usepackage[dvipsnames]{xcolor}
\usepackage{fp, tikz, pgfplots}
\usepackage{soul}
\usepackage{algorithmic}
\usepackage{algorithm}
\usepackage{booktabs} 
\usepackage{tabularx}
\usetikzlibrary{arrows,shapes,backgrounds,patterns,fadings,matrix,arrows,calc,
	intersections,decorations.markings,
	positioning,external,arrows.meta}
\tikzset{
	block/.style = {draw, rectangle,
		minimum height=1cm,
		minimum width=2cm},
	input/.style = {coordinate,node distance=1cm},
	output/.style = {coordinate,node distance=6cm},
	arrow/.style={draw, -latex,node distance=2cm},
	pinstyle/.style = {pin edge={latex-, black,node distance=2cm}},
	sum/.style = {draw, circle, node distance=1cm},
}
\usepgfplotslibrary{fillbetween}
\definecolor{nicegreen}{RGB}{0,200,0}

\newtheorem{definition}{Definition}
\newtheorem{assumption}{Assumption}
\newtheorem{remark}{Remark}
\newtheorem{theorem}{Theorem}
\newtheorem{lemma}{Lemma}

\makeatletter
\renewcommand*{\@fnsymbol}[1]{\ensuremath{\ifcase#1\or \or \dagger\or \ddagger\or
		\mathsection\or \mathparagraph\or \|\or **\or \dagger\dagger
		\or \ddagger\ddagger \else\@ctrerr\fi}}
\makeatother
\pgfplotsset{compat=1.17}
\begin{document}
\begin{frontmatter}

\title{Safe Learning-Based Control of Elastic Joint Robots via Control Barrier Functions} 

\author[]{Armin Lederer$^{\dagger}$,} 
\author[]{Azra Begzadi\'c$^{\dagger}$,} 
\author[]{Neha Das,}
\author[]{Sandra Hirche}

\address[tum]{Chair of Information-oriented Control (ITR), School of Computation, Information and Technology, Technical University of Munich, Munich, Germany (e-mail: \{armin.lederer, azra.begzadic,  neha.das, hirche\}@tum.de).}

\begin{abstract}                
Ensuring safety is of paramount importance in physical human-robot interaction applications. This requires both adherence to safety constraints defined on the system state, as well as guaranteeing compliant behavior of the robot. If the underlying dynamical system is known exactly, the former can be addressed with the help of control barrier functions. The incorporation of elastic actuators in the robot's mechanical design can address the latter requirement. However, this elasticity can increase the complexity of the resulting system, leading to unmodeled dynamics, such that control barrier functions cannot directly ensure safety. In this paper, we mitigate this issue by learning the unknown dynamics using Gaussian process regression. By employing the model in a feedback linearizing control law, the safety conditions resulting from control barrier functions can be robustified to take into account model errors, while remaining feasible. In order to enforce them on-line, we formulate the derived safety conditions in the form of a second-order cone program. We demonstrate our proposed approach with simulations on a two-degree-of-freedom planar robot with elastic joints.
\end{abstract}


\end{frontmatter}

\renewcommand*{\thefootnote}{\fnsymbol{footnote}}
\footnotetext{This work was supported by the European Research Council (ERC) Consolidator Grant 
"Safe data-driven control for human-centric systems (CO-MAN)" under grant 
agreement number 864686, by the Horizon 2020 research and innovation programme of the European Union under grant agreement number 871767 of the project ReHyb: Rehabilitation based on hybrid neuroprosthesis, and by TUM AGENDA 2030, funded by the Federal Ministry of Education and Research (BMBF) and the Free State of Bavaria under the Excellence Strategy of the Federal Government and the L\"ander as well as by the Hightech Agenda Bavaria.
}
\footnotetext[2]{Both authors contributed equally.}
\renewcommand*{\thefootnote}{\arabic{footnote}}

\section{Introduction}

Applications focusing on human-robot interactions, such as rehabilitation robotics, are highly safety-critical. The safety constraints manifest in two different ways. On the one hand, the system must satisfy state constraints to avoid damages due to the joint limits of the robot. On the other hand, the safety of humans requires the absence of peaks in the interaction forces, which has led to the application of robots with elastic joints in rehabilitation robotics to ensure compliant behavior \citep{Yu2015}.

Elasticity in the joints is commonly achieved by adding a spring between the motor and load side of a joint, which is commonly referred to as a series elastic actuator \citep{Spong1987ModelingAC}. This approach allows modeling a robot with elastic joints as a coupled system with a relative degree of four, such that a feedback linearizing controller can be straightforwardly derived \citep{Moberg2008OnFL}. Moreover, control barrier functions (CBF) \citep{Ames2017} for systems with a higher order relative degree can be easily constructed \citep{HighRelDegree}, which provides an intuitive approach to enforce state constraints on robots with series elastic actuators \citep{Nguyen2016}.

While control barrier functions are a theoretically appealing method for ensuring safety, they crucially rely on the availability of accurate models of the system dynamics. This is a particularly challenging requirement for elastic joint robots due to their comparatively high complexity. In order to mitigate this issue, supervised machine learning techniques are increasingly applied to infer models of nonlinear dynamical systems from data. In particular, Gaussian process (GP) regression is commonly employed in safety-critical applications due to its strong theoretical foundations \citep{Rasmussen2006}. When a learned model of the dynamics is used together with CBFs, the possible learning error must be taken into account to ensure the safety of the unknown system \citep{Cheng2019EndtoEndSR}. Thereby, it is possible to prove that the satisfaction of robustified CBF conditions ensures the safety of systems with learned higher relative degree dynamics with high probability \citep{Dhiman2021}. However, the feasibility of these conditions can generally not be guaranteed \citep{Castaeda2021PointwiseFO}, such that safety cannot be directly enforced using on-line optimization of the control inputs. This problem is also not avoided when learning the CBF conditions instead of the system dynamics using GPs similarly as in \citep{Greeff2021LearningAS}.

We address this lack of feasibility guarantees by proposing a novel approach for ensuring the safe control of unknown elastic joint robots via CBFs when models learned via GP regression are available. 
For this purpose, we switch between a feedback linearizing control law based on the learned model and one relying on bounds for the inertia and stiffness matrices. This allows us to exploit high probability learning error bounds to admit an effective robustification of the CBF conditions for ensuring the satisfaction of state constraints, while the matrix bounds serve as a conservative back-up to guarantee the feasibility of CBF conditions.
In order to admit the efficient enforcement of safety using on-line optimization, we reformulate the CBF conditions into second-order cone constraints.
We demonstrate the effectiveness of the proposed approach in simulations of an elastic joint robot with two degrees of freedom.\looseness=-1

The remainder of this paper is structured as follows. In Sec.~\ref{sec:problem}, we introduce elastic joint robots and formalize our problem setting. The approach for learning a model using GP regression is explained in Sec.~\ref{sec:GP}, before our approach for ensuring safety of elastic joint robots using CBFs and GP models is derived in Sec.~\ref{sec:CBF}. In Sec.~\ref{sec:Eval}, the approach is evaluated in simulations of a robot with two degrees of freedom, before the paper is concluded in Sec.~\ref{sec:conc}. 

\section{Problem Statement}\label{sec:problem}

We consider a rigid link robot with $m$ elastic joints described by differential equations\footnote{
Lower/upper case bold symbols denote vectors/matrices, $\mathbb{R}_{+, 0} / \mathbb{R}_{+} $ all real positive numbers with/without zero, $\bm{I}_m$ denotes the $m \times m$ identity matrix, $\bar{\sigma}(\cdot)/\underline{\sigma}(\cdot)$ the maximal/minimal singular values of a matrix, 
$\left\| \cdot \right\|$ the Euclidean norm, $\mathcal{N}(\mu, \sigma)$ a Gaussian distribution with mean $\mu$ and variance $\sigma$, and $\succeq$ defines the Loewner order of positive semi-definite matrices.
We denote a continuous function $\alpha: \mathbb{R}_{+} \rightarrow \mathbb{R}_{+}$ as extended class $\mathcal{K}$ function if it is strictly increasing, $\alpha(0)=0$, $\lim _{r \rightarrow \infty} \alpha(r)=\infty$ and $\lim _{r \rightarrow-\infty} \alpha(r)=-\infty$.
}\citep{Spong1987ModelingAC}
\begin{subequations}\label{eq:real_model}
\begin{align}\label{eq:real_modela}
    \bm{M}(\bm{q}) \bm{\ddot{q}}+\bm{C}(\bm{q, \dot{q}}) +\bm{K}(\bm{q}-\bm{\theta})  &=\bm{0}, \\
    \bm{J} \bm{\ddot{\theta}}+\bm{K}(\bm{\theta}-\bm{q})&= \bm{u},
    \label{eq:real_modelb}
\end{align}
\end{subequations}
where $\bm{q} \!\in\! \mathbb{R}^{m}$ represents the joint angles, $\bm{\theta} \!\in\! \mathbb{R}^{m}$ represents the motor angles, $\bm{M}(\bm{q}) \!\in\! \mathbb{R}^{m \times m}$ is the inertia matrix of the rigid links, $\bm{J} \!\in\! \mathbb{R}^{m \times m}$ is the inertia matrix of the motors, $\bm{C}(\bm{q, \dot{q}}) \!\in\! \mathbb{R}^{m}$ represents Coriolis, centrifugal and gravitational terms, $\bm{K} \!\in\! \mathbb{R}^{m \times m}$ is the matrix of stiffness coefficients, and $\bm{u} \!\in\! \mathbb{R}^{m}$ is the column vector of torque inputs provided by the motors. For the purpose of controller design and analysis, we require the following assumptions.\looseness=-1

\begin{assumption}\label{as:boundednessproperty}
    The symmetric inertia matrices $\bm{M}(\bm{q})$ and $\bm{J}$ and the stiffness matrix $\bm{K}$ are bounded above and below, i.e., there exist constants $\underline{\gamma}_{\bm{M}}, \bar{\gamma}_{\bm{M}}, \underline{\gamma}_{\bm{J}}, \bar{\gamma}_{\bm{J}}, \underline{\gamma}_{\bm{K}}, \bar{\gamma}_{\bm{K}}\in\mathbb{R}_+$ such that for all $\bm{q}\in\mathbb{R}^m$
    \begin{align}
         \underline{\gamma}_{\bm{M}} \bm{I}_m \preceq &\bm{M}(\bm{q}) \preceq \bar{\gamma}_{\bm{M}} \bm{I}_m, \\
         \underline{\gamma}_{\bm{J}} \bm{I}_m\preceq &\bm{J} \preceq \bar{\gamma}_{\bm{J}} \bm{I}_m,\\
         \underline{\gamma}_{\bm{K}} \bm{I}_m \preceq& \bm{K} \preceq \bar{\gamma}_{\bm{K}}  \bm{I}_m.
         \label{eq: kbounds}
    \end{align}
\end{assumption}

\begin{assumption}\label{as:Lipschitz}
    The functions $\bm{M}(\cdot)$ and $\bm{C}(\cdot,\cdot)$ have continuous partial derivatives up to the third order.
\end{assumption}

Ass.~\ref{as:boundednessproperty} is needed to guarantee the global controllability of the dynamics \eqref{eq:real_model} since it eliminates the possibility of internal dynamics. As it holds for all robot manipulators with only revolute or prismatic joints \citep{Ghorbel}, Ass.~\ref{as:boundednessproperty} is not restrictive in practice.  Ass.~\ref{as:Lipschitz} ensures that the functions $\bm{M}(\cdot)$ and $\bm{C}(\cdot,\cdot)$ are well-behaved, which is commonly assumed for the control design of nonlinear systems. Since the dynamics \eqref{eq:real_model} follow from an Euler-Lagrange formalism, the functions $\bm{M}(\cdot)$ and $\bm{C}(\cdot,\cdot)$ usually exhibit this required smoothness. Therefore, Ass.~\ref{as:Lipschitz} is generally not restrictive.

Since the precise identification of the parameters of robots 
with elastic actuators is a challenging problem, we merely assume that an approximate model 
\begin{subequations}\label{eq:approx_model}
\begin{align}
    \hat{\bm{M}}(\bm{q}) \bm{\ddot{q}}+\hat{\bm{C}}(\bm{q, \dot{q}}) +\hat{\bm{K}}(\bm{q}-\bm{\theta})  &=\bm{0} \\
    \hat{\bm{J}} \bm{\ddot{\theta}}+\hat{\bm{K}}(\bm{\theta}-\bm{q})&= \bm{u}
\end{align}
\end{subequations}
is known, while the true dynamics \eqref{eq:real_model} are unknown. In order to infer a model of the residual error between the true system and the approximate model, we consider the availability of training data as described in the following.
\begin{assumption}
\label{as:data}
A data set
\begin{align}
    \mathbb{D}=\Big\{ \bm{q}^{(n)},\dot{\bm{q}}^{(n)},\ddot{\bm{q}}^{(n)},\overset{( 3 )}{\dot{\bm{q}}}{}^{(n)}, \overset{(4)}{\dot{\bm{q}}}{}^{(n)}+\bm{\omega}^{(n)} \Big\}_{n=1}^{N}
\end{align}
is available, which contains $N$ quintuples consisting of noise-free measurements of joint angles and their derivatives, while the fourth order derivatives are
perturbed by Gaussian noise $\omega^{(n)} \sim \mathcal{N}\left(0, \sigma_{\mathrm{on}}^{2}\right)$.
\end{assumption}

The assumption that only the highest derivative of a signal is perturbed by Gaussian noise can commonly be found in the literature when inferring models of unknown dynamics \citep{Lederer2020c,Dhiman2021, Greeff2021LearningAS}. While this might be difficult to achieve in practice, numerical differentiation approaches ensure that noise in lower order derivatives is comparatively small. Since the focus of this paper is on the development of a safe control approach for elastic joint robots with learned models, we leave the extension to training data sets where all samples are perturbed by noise to future work.

Based on these assumptions, we consider the problem of designing a control law $\bm{\pi}:\mathbb{R}^m\times\mathbb{R}^m\times\mathbb{R}^m\times\mathbb{R}^m\rightarrow\mathbb{R}^m$ which ensures the safety of the robotic system with elastic joints. In this paper, we examine safety in terms of state constraints expressed through the zero-super level set
\begin{align}\label{eq:const_set}
    \mathcal{C} =\left\{\bm{q} \in \mathbb{R}^m\!: b(\bm{q}) \geq 0\right\}, &&\partial\mathcal{C} = \left\{\bm{q} \in \mathbb{R}^m\!: b(\bm{q}) = 0\right\}\!
\end{align}
of an arbitrary function $b: \mathbb{R}^m \rightarrow \mathbb{R}$ with continuous derivatives up to the fourth order. Therefore, safety essentially reduces to forward invariance of $\mathcal{C}$, as formalized in the following.
\begin{definition}[Safety~\citep{Ames2017}]
A system \eqref{eq:real_model}  is safe with respect to the set $\mathcal{C}$ if the set $\mathcal{C}$ is forward invariant, i.e., for any initial condition $\bm{x}_0 \in \mathcal{C}$, it holds that $\bm{x}(t) \in \mathcal{C}$  for $\bm{x}(0)=\bm{x}_0 $ and all $t\geq 0$.
\end{definition}

\section{Learning Gaussian Process Models of Control-Affine Systems}\label{sec:GP}

In order to learn a model of elastic joint robots, we employ GP regression \citep{Rasmussen2006}. The fundamentals of GP regression are explained in Sec.~\ref{subsec:GP}, before we show how control-affine models with error bounds can be learned in Sec.~\ref{subsec:ctrl_aff_learn}.

\subsection{Gaussian Process Regression}\label{subsec:GP}

Gaussian process regression is a supervised machine learning method, which relies on the assumption that any finite number of evaluations $\{h(\bm{x}^{(1)}),\ldots,h(\bm{x}^{(N)})\}$, $N\in\mathbb{N}$, of an unknown function $h:\mathbb{R}^d\rightarrow\mathbb{R}$ at inputs $\bm{x}\in\mathbb{R}^d$ follow a joint Gaussian distribution. A Gaussian process, denoted as $\mathcal{GP}(\hat{h}(\cdot),k(\cdot,\cdot))$ is fully specified using a prior mean $\hat{h}:\mathbb{R}^d\rightarrow\mathbb{R}$ and a covariance function $k:\mathbb{R}^d\times\mathbb{R}^d\rightarrow\mathbb{R}_+$. The mean function incorporates prior model knowledge in the form of an approximate model into the regression, while the covariance function encodes abstract information about the structure of the regressed function such as differentiability. 

When training data $\{\bm{x}^{(n)},y^{(n)}\}_{n=1}^N$ with Gaussian perturbed training targets $y^{(n)}=h(\bm{x}^{(n)})+\epsilon^{(n)}$, $\epsilon^{(n)}\sim\mathcal{N}(0,\sigma_{\mathrm{on}}^2)$ is available, the joint Gaussian distribution of function evaluations can straightforwardly be exploited to perform regression by determining the posterior distribution. Due to the properties of Gaussian random variables, this distribution is again Gaussian with mean and variance 
\begin{align}
\mu(\bm{x}) &=\hat{h}(\bm{x})+\bm{k}^{T}(\bm{x})\left(\bm{K}+\sigma_{\mathrm{on }}^{2} \bm{I}_{N}\right)^{-1} (\bm{y}-\hat{\bm{h}}), \\
\sigma^{2}(\bm{x}) &=k(\bm{x}, \bm{x})-\bm{k}^{T}(\bm{x})\left(\bm{K}+\sigma_{\mathrm{on}}^{2} \bm{I}_{N}\right)^{-1} \bm{k}(\bm{x}),
\end{align}
where $\bm{k}(\bm{x})$ and $\bm{K}$ are defined element-wise via $k_i(\bm{x})=k(\bm{x},\bm{x}^{(i)})$ and $K_{ij}=k(\bm{x}^{(i)},\bm{x}^{(j)})$, respectively, $\hat{\bm{h}}=[\hat{h}(\bm{x}^{(1)})\ \cdots\ \hat{h}(\bm{x}^{(N)})]^T$, and $\bm{y}=[y^{(1)}\ \cdots\ y^{(N)}]^T$.

\subsection{Learning Models of Control-Affine Systems}\label{subsec:ctrl_aff_learn}

While Gaussian process regression is often employed for completely unknown functions $h(\cdot)$, for efficient control design, we often know the types of function structure the model must adhere to. A very common structure makes the dynamical system affine to the control input, which yields training targets of the form
\begin{align}\label{eq:aff_dyn}
    \bm{y}=\bm{h}(\bm{x},\bm{u})+\bm{\omega}=\bm{f}(\bm{x})+\bm{G}(\bm{x})\bm{u}+\bm{\omega},
\end{align}
where $\bm{f}\!:\mathbb{R}^m\!\rightarrow\!\mathbb{R}^m$, $\bm{G}\!:\mathbb{R}^{m}\!\rightarrow\!\mathbb{R}^m\!\times\mathbb{R}^m$ are unknown functions and $\bm{\omega}\!\sim\!\mathcal{N}(\bm{0},\sigma_{\mathrm{on}}^2\bm{I}_m)$ is Gaussian observation noise. In order to encode this structure into regression, we put a GP prior on each individual element of $\bm{f}(\cdot)$ and $\bm{G}(\cdot)$, i.e.,\looseness=-1
\begin{align}\label{eq:fprior}
    f_{i}(\cdot)&\sim \mathcal{G P}\left(\hat{f}_{i}(\cdot), k_{f_i}\left(\cdot, \cdot\right)\right),\quad i=1,\ldots,m,\\
    g_{ij}(\cdot)&\sim \mathcal{G P}\left(\hat{g}_{ij}(\cdot), k_{g_{ij}}\left(\cdot, \cdot\right)\right),\quad i,j=1,\ldots,m.
    \label{eq:gprior}
\end{align}

This implies for each row of \eqref{eq:aff_dyn} that
\begin{align}
    f_i(\bm{x})\!+\!\sum\limits_{j=1}^m g_{ij}(\bm{x})u_j\sim\mathcal{GP}(&\hat{h}_i(\bm{x},\bm{u}), k_i(\bm{x},\bm{u},\bm{x}',\bm{u}')),
\end{align}
where we have the composite means and kernels
\begin{align}
    \hat{h}_i(\bm{x},\bm{u})&=\hat{f}_i(\bm{x})\!+\!\sum\limits_{j=1}^m \hat{g}_{ij}(\bm{x})u_j,\\
    k_i(\bm{x},\bm{u},\bm{x}',\bm{u}')&=k_{f_i}(\bm{x},\bm{x}')\!+\!\sum\limits_{j=1}^m u_j k_{g_{ij}}(\bm{x},\bm{x}')u_j'.
\end{align}
Using these priors, it is straightforward to derive the posterior distributions of functions $f_i(\cdot)$ and $g_{ij}(\cdot)$ analogously to standard GP regression by conditioning the joint prior of the individual functions $f_i(\cdot)$/$g_{ij}(\cdot)$ and $h_i(\cdot)$ on the training data \citep{Duvenaud2014}.  The resulting posteriors are again Gaussian with means
\begin{align}
\mu_{f_i}(\bm{x})&=\hat{f_i}(\bm{x})+\bm{k}_{f_i}^{T}(\bm{x})\left(\bm{K}_i+\sigma_{\mathrm{on}}^{2} \bm{I}_{N}\right)^{-1} \tilde{\bm{y}}_i, \\
\mu_{g_{ij}}(\bm{x})&=\hat{g}_{ij}(\bm{x})+\bm{k}_{g_{ij}}^{T}(\bm{x}) \bm{U}_j\left(\bm{K}_i+\sigma_{\mathrm{on}}^{2} \bm{I}_{N}\right)^{-1} \tilde{\bm{y}}_i 
\end{align}
and variances
\begin{align}
    \sigma_{f_i}^{2}(\bm{x})&=k_{f_i}(\bm{x}, \bm{x})\nonumber\\
&\quad~ -\bm{k}_{f_i}^{T}(\bm{x})\left(\bm{K}_i+\sigma_{\mathrm{on}}^{2} \bm{I}_{N}\right)^{-1} \bm{k}_{f_i}(\bm{x}), \\
\sigma_{g_{ij}}^{2}(\bm{x})&=k_{g_{ij}}(\bm{x}, \bm{x}) \nonumber\\
&\quad~ -\bm{k}_{g_{ij}}^{T}(\bm{x}) \bm{U}_{j}\left(\bm{K}_i+\sigma_{\mathrm{on}}^{2} \bm{I}_{N}\right)^{-1} \bm{U}_j \bm{k}_{g_{ij}}(\bm{x}),\!
\end{align}
where $\tilde{y}_{i}^{(n)}=y_{i}^{(n)}-\hat{f}_i(\bm{x}^{(n)})-\sum_{j=1}^{m}\hat{g}_{ij}(\bm{x}^{(n)})u_j^{(n)} $, $\bm{U}_j=\operatorname{diag}([u_j^{(1)} \ldots u_j^{(N)}])$ and $\bm{K}_i =\bm{K}_{f_i}+\sum_{j=1}^{m}\bm{U}_j \bm{K}_{g_{i,j}} \bm{U}_j$.

Due to the strong theoretical foundations of Gaussian process regression, it is straightforward to extend Bayesian prediction error bounds \citep{Lederer2019UniformEB} to the individual learned functions as shown in the following lemma.
\begin{lemma}\label{lem:GPbound}
Assume that the functions $f_i(\cdot)$, $g_{ij}(\cdot)$ are sample functions from corresponding GPs, i.e., \eqref{eq:fprior} and \eqref{eq:gprior} hold. Then, there exists a constant $\beta\in\mathbb{R}_{+}$ and a probability $\delta\in (0,1)$ such that 
\begin{align}\label{eq:GPboundf}
    P\left(|\mu_{f_i}(\bm{x})\!-\!f_i(\bm{x})|\leq \sqrt{\beta}\sigma_{f_i}(\bm{x})~\forall\bm{x}\in\mathbb{X}  \right)&\geq 1\!-\!\delta\\
    P\left(|\mu_{g_{ij}}(\bm{x})\!-\!g_{ij}(\bm{x})|\leq \sqrt{\beta}\sigma_{g_{ij}}(\bm{x})~\forall\bm{x}\in\mathbb{X}  \right)&\geq 1\!-\!\delta
    \label{eq:GPboundg}
\end{align}
holds for a compact set $\mathbb{X}\subset\mathbb{R}^d$.
\end{lemma}
\begin{pf}
The proof straightforwardly follows by extending \citep[Lemma 1]{Lederer2021HowTD} to multiple summands and in combination with the choice of a sufficiently large value for $\beta$ \citep[Proposition 1]{Lederer2022}. \hfill$\square$
\end{pf}
While this lemma ensures only the existence of a constant $\beta$, this limitation is used merely for notational simplicity. It is straightforward to compute a value $\beta$ in practice using the results in \citep{Lederer2021HowTD, Lederer2022}. Therefore, this result enables the quantification of the possible model error, which we use for the robustification of safety conditions.
\section{Safe Control of Elastic Joint Robots using Gaussian Process Models}\label{sec:CBF}

\begin{figure}
\begin{center}
\begin{tikzpicture}[auto, node distance=0.5cm,>=latex', scale=0.66, transform shape] 
	\node [input, name=input] {};
	\node [sum, right=1.5cm of input, line width=0.25mm,] (sum) {};
	
	\node [draw=black,
	minimum width=2cm,
	minimum height=1.2cm,
	right=0.5cm of sum]   (controller) {\begin{tabular}{c} tracking \\ controller \end{tabular}};
	\node [draw=black,
	minimum width=2cm,
	minimum height=1.2cm,
	right=0.5cm of controller]  (plant) {CBF-SOCP};        
	
	\node [draw=black,
	minimum width=2cm,
	minimum height=1.2cm, 
	right=0.5cm of plant]  (cbf) {\begin{tabular}{c} feedback \\ linearization \end{tabular}};
         
	\node [draw=black,
	minimum width=2cm,
	minimum height=1.2cm,
	right=0.5cm of cbf]  (plant1){\begin{tabular}{c} elastic \\ joint robot \end{tabular}};    	
	\node [draw=black,
	minimum width=2cm,
	minimum height=1.2cm,
	above=0.3cm of plant1]  (gp){GP model}; 
	
	\node [draw=black,
	minimum width=2cm,
	minimum height=1.2cm,
	above=0.3cm of gp]  (backup){prior model}; 
	
	\node [draw=black,
	minimum width=2cm,
	minimum height=1.2cm,
	above=1.055cm of cbf]  (switch){}; 
	
	\node [minimum width=1mm,
	line width=0.25mm,
	minimum height=1.4cm, output, right=0.3cm of plant1] (y) {};
	
	\draw [draw,->, line width=0.25mm] (input) -- node {reference} (sum);
	\draw [->,line width=0.25mm] (sum) -- (controller);
	\draw [->,line width=0.25mm] (controller) -- (plant);
	\draw [->,line width=0.25mm] (plant) -- (cbf);
	\draw [->,line width=0.25mm] (cbf) --  (plant1);
	\draw [line width=0.25mm] (plant1) --  (y);
	\draw [->, line width=0.25mm] (y) -- ++ (0,-0.9) -| node [pos=0.98, right] {$-$} (sum.south);
	\draw [->, line width=0.25mm] (y) -- ++ (0,-0.9)  |- node [pos=0.98, left] {} (gp.east);
	\draw [->,line width=0.25mm] (y) -- ++ (0,-0.9) -| node [pos=0.98, left] {} (plant);
	\draw [->,line width=0.25mm] (y)  |- node [pos=0.98, left] {} (backup.east);
	
	\draw [line width=0.25mm] (switch.east)+(-5mm,4mm) -| ++ (0.4,0.76) -- (backup.west);
	\draw [line width=0.25mm] (switch.east)+(-5mm,-4mm) -| ++ (0.4,-0.755) -- (gp.west);
	\draw[line width=0.5mm] (switch.east)+(-6mm,4mm) circle (0.1cm);
	\draw[line width=0.5mm] (switch.east)+(-6mm,-4mm) circle (0.1cm);
	\draw[line width=0.5mm] (switch.east)+(-5.8mm,-2.8mm) -- ++ (-15mm,0mm);
	\draw[line width=0.25mm] (switch.east)+(-15mm,0mm) -- (switch.west);
	\draw[line width=0.5mm,fill=black] (switch.east)+(-15mm,0mm) circle (0.1cm);
	\draw[line width=0.25mm,->] (switch.east)+(-6mm,-2.1mm) arc (-13.5:13.5:1);
	
	\draw[line width=0.25mm,->] (switch.west) -| ++(-5.3mm,-10mm) -| (cbf.north); 
	\draw[line width=0.25mm,->] (switch.west)+(-5.3mm,-10mm) -| (plant.north); 
\end{tikzpicture}\vspace{-0.2cm}
  \caption{
  Safety of elastic joint robots is ensured using control barrier functions derived for a switching system, which results from a feedback linearization using a learned model or a conservative prior model. 
  }
        \label{fig:1}
    \end{center}
    \end{figure}
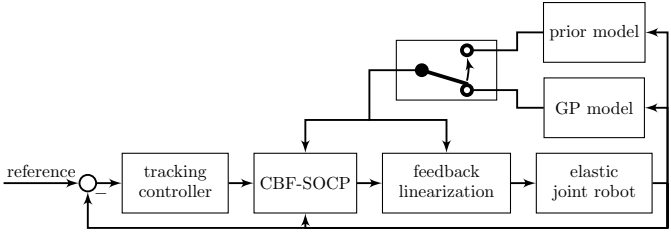

Since the learned model of the elastic joint robot exhibits model errors, we need to employ robust CBF conditions, which potentially can be infeasible. As outlined in Fig.~\ref{fig:1}, we approach this issue using a feedback linearizing controller which switches between the GP model and a back-up model based on conservative model error bounds. Bounds for the linearization errors of these control laws are presented in Sec.~\ref{subsec:FeLi}. In Sec.~\ref{subsec:CBF}, robust CBF conditions for ensuring the safety of unknown elastic joint robots are derived and a switching strategy for ensuring their feasibility on-line is developed. By reformulating the CBF conditions into a second\textcolor{blue}{-}order cone program in Sec.~\ref{subsec:SOCP}, we provide an efficient method for enforcing safety of arbitrary control laws on-line.

\subsection{Feedback Linearization for Elastic Joint Robots}\label{subsec:FeLi}

In order to develop a safe controller for elastic joint robots, we follow the idea of \citet{Moberg2008OnFL} and reformulate the dynamics, such that they admit a feedback linearization. Therefore, we sum \eqref{eq:real_modela} and \eqref{eq:real_modelb} yielding
\begin{align}
    \bm{M}(\bm{q}) \bm{\ddot{q}}+\bm{J} \bm{\ddot{\theta}}+
    \bm{C}(\bm{q, \dot{q}})&=\bm{u}.
\end{align}
We solve this equation for $\bm{\theta}$ and substitute the result into \eqref{eq:real_modela} after differentiating it twice. This allows us to express the dynamics as the control-affine system 
\begin{align}\label{eq:ctrl_aff_sys}
    \dot{\bm{x}}_1\!=\!\bm{x}_2,&& \dot{\bm{x}}_2\!=\!\bm{x}_3,&& \dot{\bm{x}}_3\!=\!\bm{x}_4, && \dot{\bm{x}}_4\!=\!\bm{f}(\bm{x})+\bm{G}(\bm{x})\bm{u},
\end{align}
where
\begin{align}\label{eq:f_FeLi}
\bm{f}(\bm{x})=&-\bm{M}^{-1}\left(\bm{x}_{1}\right) \Big(\bm{K} \bm{J}^{-1}\left(\bm{M}\left(\bm{x}_{1}\right) \bm{x}_{3}+\bm{C}\big(\bm{x}_{1}, \bm{x}_{2}\right)\big) \nonumber \\
& -\big(\bm{K}+\bm{\ddot{M}}\left(\bm{x}_{1}, \bm{x}_{2}, \bm{x}_{3}\right)\big) \bm{x}_{3}- 2\bm{\dot{M}}\left(\bm{x}_{1}, \bm{x}_{2}\right) \bm{x}_{4}\nonumber \\
& - \bm{\ddot{C}}\left(\bm{x}_{1}, \bm{x}_{2}, \bm{x}_{3}, \bm{x}_{4}\right)\Big), \\
\bm{G}(\bm{x})=& \bm{M}^{-1}\left(\bm{x}_{1}\right) \bm{K} \bm{J}^{-1} , \label{eq:g_FeLi}
\end{align} 
with  $\bm{x}_1=\bm{q}$. Due to the structure of \eqref{eq:ctrl_aff_sys}, the states $\bm{x}_i$ correspond to the joint angles and their derivatives. We concatenate them into a vector, i.e., $\bm{x}=[\bm{x}_1^T\ \bm{x}_2^T\ \bm{x}_3^T\ \bm{x}_4^T]^T$, such that 
we can employ the training data from Ass.~\ref{as:data} to train a Gaussian process model as explained in Sec.~\ref{subsec:ctrl_aff_learn}. Note that we can include the prior knowledge of $\bm{M}(\cdot)$, $\bm{C}(\cdot,\cdot)$, $\bm{K}$ and $\bm{J}$ via suitable prior mean functions $\hat{f}_i(\cdot)$ and $\hat{g}_{ij}(\cdot)$ reflecting the structure of \eqref{eq:f_FeLi} and \eqref{eq:g_FeLi}, respectively. In order to employ the result of GP regression in a control law, we concatenate the elements $\mu_{f_i}(\cdot)$ into a vector $\bm{\mu}_f(\cdot)$ and the elements $\mu_{g_{ij}}(\cdot)$ into a matrix $\bm{\mu}_{G}(\cdot)$.
Then, we can define the feedback linearizing control law
\begin{align}\label{eq:ctrl_GP}
    \bm{\pi}_{GP}(\bm{x})=\bm{\mu}_G^{-1}(\bm{x})(\bm{\nu}-\bm{\mu}_f(\bm{x})),
\end{align}
where $\bm{\nu}$ is the control input to the approximately linearized system under the assumption that $\bm{\mu}_{G}(\cdot)$ is invertible.
Using this control law, we can compactly express the controlled system as
\begin{align}\label{eq:FeLi_dyn}
    \dot{\bm{x}}=\bm{A}\bm{x}+\bm{B}\left(\bm{\nu}+\bm{e}_{GP}(\bm{x})+\bm{E}_{GP}(\bm{x})\bm{\nu}\right),
\end{align}
where the dynamic parameters are given by
\begin{align}
    \bm{A}=\begin{bmatrix} \bm{0}&\bm{I}_m&\bm{0}&\bm{0}\\\bm{0}&\bm{0}&\bm{I}_m&\bm{0}\\\bm{0}&\bm{0}&\bm{0}&\bm{I}_m\\\bm{0}&\bm{0}&\bm{0}&\bm{0}
    \end{bmatrix},&&
    \bm{B}=\begin{bmatrix}\bm{0}\\\bm{0}\\\bm{0}\\\bm{I}_m
    \end{bmatrix},
\end{align}
and the linearization errors due to using a learned model are denoted by
\begin{align}
    \bm{e}_{GP}(\bm{x}) &=
    \bm{f}(\bm{x})-\bm{\mu}_f(\bm{x})-\bm{E}_{GP}(\bm{x})\bm{\mu}_f(\bm{x}), \\ 
    \bm{E}_{GP}(\bm{x}) &= (\bm{G}(\bm{x})-\bm{\mu}_G(\bm{x}))\bm{\mu}_G^{-1}(\bm{x}).
\end{align}
Due to the definition of $\bm{E}_{GP}(\bm{x})$, its maximum singular value is bounded by
\begin{align}
    \|\bm{E}_{GP}(\bm{x})\|\bar{\sigma}(\bm{E}_{GP}(\bm{x}))\leq \frac{\underline{\sigma}(\bm{G}(\bm{x})-\bm{\mu}_{\bm{G}}(\bm{x}))}{\underline{\sigma}(\bm{\mu}_{\bm{G}}(\bm{x}))}.
\end{align}
While we cannot evaluate this inequality directly due to a lack of knowledge of $\bm{G}(\cdot)$, we can employ the GP prediction error bound in Lemma~\ref{lem:GPbound} to obtain 
\begin{align}
    \underline{\sigma}(\bm{G}(\bm{x})\!-\!\bm{\mu}_{\bm{G}}(\bm{x}))&\leq \|\bm{G}(\bm{x})\!-\!\bm{\mu}_{\bm{G}}(\bm{x})\|_{\mathrm{Fr}}\\
    &=\sqrt{\beta \sum\limits_{i=1}^m\sum\limits_{j=1}^m \sigma_{g_{ij}}^2(\bm{x})}
\end{align}
with probability of at least $1-m^2\delta$, where $\|\cdot\|_{\mathrm{Fr}}$ denotes the Frobenius norm. Therefore, we have 
\begin{align}\label{eq:sigma_EGP}
    \bar{\sigma}(\bm{E}_{GP}(\bm{x}))
    \!\leq\!  \gamma_{\bm{E}_{GP}}(\bm{x})=\frac{\sqrt{\beta \sum\limits_{i=1}^m\sum\limits_{j=1}^m \sigma_{g_{ij}}^2(\bm{x})}}{\underline{\sigma}(\bm{\mu}_G(\bm{x}))},
\end{align}
and consequently,
\begin{align}\label{eq:e_GP}
    \|\bm{e}_{GP}(\bm{x})\|\leq\bar{e}_{GP}= \sqrt{\beta}\|\bm{\sigma}_f(\bm{x})\| \!+\! \gamma_{\bm{E}_{GP}}(\bm{x}) \|\bm{\mu}_f(\bm{x})\|
\end{align}
with probability of at least $1-(m^2+m)\delta$, where $\bm{\sigma}_f(\bm{x})=[\sigma_{f_1}(\bm{x})\ \cdots\ \sigma_{f_m}(\bm{x})]^T$.

In order to obtain a conservative, back-up model for feedback linearization, we exploit Ass.~\ref{as:boundednessproperty} to derive bounds for $\bm{G}(\bm{x})$ in the Loewner order as shown in the following.

\begin{lemma}
Under Ass.~\ref{as:boundednessproperty}, the matrix $\bm{G}(\bm{x})$ is positive definite and 
it holds that
    \begin{equation}\label{eq:Gbound}
    \underline{\gamma}_{\bm{G}} \bm{I} \preceq \bm{G}(\bm{x}) \preceq \bar{\gamma}_{\bm{G}} \bm{I}, \quad \forall \bm{x} \in \mathbb{R}^{n} ,
\end{equation}
where $\underline{\gamma}_{\bm{G}} = \frac{\underline{\gamma}_{\bm{K}}}{\bar{\gamma}_{\bm{M}} \bar{\gamma}_{\bm{J}}}$ and $\bar{\gamma}_{\bm{G}} = \frac{\bar{\gamma}_{\bm{K}}}{\underline{\gamma}_{\bm{M}} \underline{\gamma}_{\bm{J}}}$. \label{as:g4bounds}
\end{lemma}
\begin{pf}
Due to the definition of $\bm{G}(\bm{x})$ in \eqref{eq:g_FeLi} and Ass.~\ref{as:boundednessproperty}, it directly follows that 
\begin{equation*}
    \frac{\underline{\gamma}_{\bm{K}}}{\bar{\gamma}_{\bm{M}} \bar{\gamma}_{\bm{J}}} \bm{I} \preceq \bm{M}^{-1}\left(\bm{x}_{1}\right) \bm{K} \bm{J}^{-1} \preceq  \frac{\bar{\gamma}_{\bm{K}}}{\underline{\gamma}_{\bm{M}} \underline{\gamma}_{\bm{J}}} \bm{I},
\end{equation*}
which concludes the proof.\hfill$\square$
\end{pf}

This lemma allows us to define the back-up control law
\begin{align}
    \bm{\pi}_{\gamma}(\bm{x})=\frac{1}{\bar{\gamma}_{\bm{G}}}(\bm{\nu}-\bm{\mu}_f(\bm{x})),
\end{align}
which leads to the linearization error 
\begin{align}
    \bm{e}_{\gamma}(\bm{x}) &= \bm{f}(\bm{x})\!-\!\bm{\mu}_f(\bm{x}) \!-\!\bm{E}_{\gamma}(\bm{x})\bm{\mu}_f(\bm{x}),\!\\
    \bm{E}_{\gamma}(\bm{x}) &= \frac{\bm{G}(\bm{x})\!-\!\bar{\gamma}_{\bm{G}}\bm{I}_m }{\bar{\gamma}_{\bm{G}}}.
    \label{eq:Egamma}
\end{align}
Due to the lower bound on $\bm{G}(\cdot)$ in \eqref{eq:Gbound}, the linearization error for this controller can directly be bounded by
\begin{align}
    \|\bm{E}_{\gamma}(\bm{x})\|=\bar{\sigma}(\bm{E}_{\gamma}(\bm{x}))
    \leq\gamma_{\bm{E}_{\gamma}}(\bm{x})= \frac{\bar{\gamma}_{\bm{G}}-\underline{\gamma}_{\bm{G}}}{\bar{\gamma}_{\bm{G}}}<1,\label{eq:sigma_Egamma}
\end{align}
and consequently
\begin{align}\label{eq:e_gamma}
    \|\bm{e}_{\gamma}(\bm{x})\|\leq\bar{e}_{\gamma}= \sqrt{\beta}\|\bm{\sigma}_f(\bm{x})\| + \gamma_{\bm{E}_{\gamma}}(\bm{x}) \|\bm{\mu}_f(\bm{x})\|.
\end{align}

\vspace{0.1cm}
\begin{remark}
It straightforwardly follows from \citep{Lederer2021HowTD} that the direct application of $\bm{\pi}_{GP}(\bm{x})$, $\bm{\pi}_{\gamma}(\bm{x})$ and any control law switching between them after a positive time ensures an ultimately bounded closed-loop system for a suitable input $\bm{\nu}$ and systems of the form \eqref{eq:ctrl_aff_sys}.
\end{remark}



\subsection{Control Barrier Functions for Learned GP Models}\label{subsec:CBF}

In order to ensure the safety of series elastic actuators with respect to the set $\mathcal{C}$, the input $\bm{\nu}$ to an approximately linearized system of the form \eqref{eq:FeLi_dyn} must render $\mathcal{C}$ forward invariant. This can be straightforwardly shown using the concept of (control) barrier functions resulting in the following lemma by \citet[Proposition 1]{Ames2017}.
\begin{lemma} \label{lem:cbf}
If there exists a continuously differentiable function $\psi:\mathbb{R}^{4m}\rightarrow\mathbb{R}$, called control barrier function (CBF), and an extended class $\mathcal{K}_{\infty}$ function $\alpha:\mathbb{R}\rightarrow\mathbb{R}$ for the approximately linearized system \eqref{eq:FeLi_dyn} controlled by a controller $\bm{\pi}_{\bm{\nu}}:\mathbb{R}^{4m}\rightarrow\mathbb{R}^m$, such that 
\begin{align}\label{eqn:cbf}
\!\nabla_{\bm{x}}^T\psi(\bm{x})( \bm{A}\bm{x}\!+\!\bm{B}((\bm{I}_m\!+\!\bm{E}_{GP}(\bm{x}))\bm{\pi}_{\bm{\nu}}(\bm{x})\!+\!\bm{e}_{GP}&(\bm{x}) )\geq\!\\
&-\alpha(\psi(\bm{x}))\nonumber
\end{align}
holds for all $\bm{x}\in\mathbb{X}$, then, the set $\mathcal{C}$ is safe.
\end{lemma}
The term $\bm{I}_m+\bm{E}_{GP}(\bm{x})$ has a crucial role in \eqref{eqn:cbf} since it determines how the controller $\bm{\pi}_{\bm{\nu}}(\cdot)$ can influence the safety of the system. This can be easily seen when considering a singular matrix $\bm{I}_m+\bm{E}_{GP}(\bm{x})$, which can prevent the existence of any control law $\bm{\pi}_{\bm{\nu}}(\cdot)$ satisfying \eqref{eqn:cbf}. Since positive singular values of $\bm{I}_m+\bm{E}_{GP}(\bm{x})$ guarantee its non-singularity, a straightforward condition for avoiding this worst case is given by $\|\bm{E}_{GP}\|<1$. Due to \eqref{eq:sigma_EGP}, this is ensured for the GP-based feedback linearizing controller \eqref{eq:ctrl_GP} if\looseness=-1
\begin{align}\label{eq:GP_cond}
    \sqrt{\beta \sum\limits_{i=1}^m\sum\limits_{j=1}^m \sigma_{g_{ij}}^2(\bm{x})}<\underline{\sigma}(\bm{\mu}_G(\bm{x})).
\end{align}
Note that the satisfaction of this inequality also guarantees the invertibility of $\bm{\mu}_G(\bm{x})$ as it implies positive singular values of $\bm{\mu}_G(\bm{x})$.
Based on these insights, we can define a switching control law 
\begin{align}\label{eq:FelLi_pol}
    \bm{\pi}(\bm{x}) = \begin{cases}
    \bm{\pi}_{GP}(\bm{x})&\text{if } \gamma_{\bm{E}_{GP}}(\bm{x})\leq \zeta\\
    \bm{\pi}_{\gamma}(\bm{x})&\text{otherwise}
    \end{cases}
\end{align}
and the corresponding linearization errors
\begin{align}
    \bm{e}(\bm{x}) = \begin{cases}
    \bm{e}_{GP}(\bm{x})&\gamma_{\bm{E}_{GP}}(\bm{x})\leq \zeta\\
    \bm{e}_{\gamma}(\bm{x})&\text{otherwise},
    \end{cases}\\
    \bm{E}(\bm{x}) = \begin{cases}
    \bm{E}_{GP}(\bm{x})&\gamma_{\bm{E}_{GP}}(\bm{x})\leq \zeta\\
    \bm{E}_{\gamma}(\bm{x})&\text{otherwise},
    \end{cases}\label{eq:E(x)}
\end{align}
where condition \eqref{eq:GP_cond} is slightly tightened using a constant $\zeta\in(0,1)$ to avoid the strict inequality.

In order to guarantee the existence of a safe control law $\bm{\pi}_{\bm{\nu}}(\cdot)$ satisfying \eqref{eqn:cbf}, it remains to design a suitable control barrier function $\psi(\cdot)$ to express constraints of the form \eqref{eq:const_set}.
For this purpose, we make use of the iterative construction proposed by \citep{HighRelDegree}, which determines a CBF by differentiating the constraint function $b(\cdot)$ according to the relative degree of the dynamics \eqref{eq:FeLi_dyn}. Defining $\tilde{\psi}_{1}(\bm{x}) = b(\bm{q})$, this leads to the following definition of the control barrier function
\begin{subequations}\label{eq:cbf_construct}
\begin{align}
    \tilde{\psi}_{i}(\bm{x}) &=\dot{\tilde{\psi}}_{i-1}(\bm{x})+\alpha_{i}\left(\tilde{\psi}_{i-1}(\bm{x})\right),\\
    \psi(\bm{x})&=\tilde{\psi}_4(\bm{x}),
\end{align}
\end{subequations}
where $\alpha_i(\cdot)$ can be simply chosen to be, e.g., identity maps.

Due to the design of the proposed switching strategy for the GP-based feedback linearizing control law \eqref{eq:FelLi_pol} and the iterative construction of the CBFs \eqref{eq:cbf_construct}, it is straightforward to show that an input $\bm{\nu}$ ensuring safety exists for every state $\bm{x}$.
\begin{theorem}\label{th:safety_existence}
Consider an elastic joint robot \eqref{eq:real_model} satisfying Ass.~\ref{as:boundednessproperty}. Assume that for its system description \eqref{eq:ctrl_aff_sys}, the functions $f_i(\cdot)$, $g_{ij}(\cdot)$ are sample functions from corresponding GPs, i.e., \eqref{eq:fprior} and \eqref{eq:gprior} hold. 
Then, given a constraint function $b(\cdot)$ that is at least $4$ times continuously differentiable, there exists a safe control law $\bm{\pi}_{\bm{\nu}}(\cdot)$ with probability of at least $1-m\delta^2$.
\end{theorem}
\begin{pf}
In order to prove this theorem, we need to show the existence of a vector $\bm{\nu}=\bm{\pi}_{\bm{\nu}}(\bm{x})$ such that \eqref{eqn:cbf} is satisfied. For this purpose, set $\bm{\nu}=\gamma \bm{B}^T\nabla_{\bm{x}}\psi(\bm{x})$. Note that only the summands $\bm{B}\bm{\nu}$ and $\bm{B}\bm{E}(\bm{x})\bm{\nu}$ depend on $\bm{\nu}$, for which we can easily see that, for $\gamma\geq 0$,  
\begin{align}\label{eq:gamma_ineq}
    \nabla_{\bm{x}}^T\psi(\bm{x})\bm{B} (\bm{I}\!+\!\bm{E}(\bm{x}))\bm{\nu}\!\geq\! \gamma (1\!-\!\|\bm{E}(\bm{x})\|) \|\nabla_{\bm{x}}^T\psi(\bm{x})\bm{B}\|^2 \!\!
\end{align}
holds. Due to the definition of $\bm{E}(\bm{x})$ in \eqref{eq:E(x)}, it follows from \eqref{eq:sigma_EGP}, \eqref{eq:sigma_Egamma}, \eqref{eq:GP_cond} and Lemma~\ref{lem:GPbound} that $\|\bm{E}(\bm{x})\|<1$ with probability of at least $1-m^2\delta$. Moreover, the construction of $\psi(\cdot)$ in \eqref{eq:cbf_construct} ensures that $\|\nabla_{\bm{x}}^T\psi(\bm{x})\bm{B}\|^2>0$ holds \citep{HighRelDegree}.  Therefore, we obtain
\begin{align*}
    \sup _{\bm{\nu}\in\mathbb{R}^m} \nabla_{\bm{x}}^T\psi(\bm{x})\left( \bm{A}\bm{x}+\bm{B}\left(\bm{\nu}+\bm{e}(\bm{x})+\bm{E}(\bm{x})\bm{\nu}\right) \right)=\infty,
\end{align*}
with probability of at least $1\!-\!m\delta^2$ for all $\bm{x}\!\in\!\mathbb{X}$, which ensures the satisfaction of \eqref{eqn:cbf}, such the lemma directly follows from Lemma~\ref{lem:cbf} and the straightforward extension to switched systems \citep{Kivilcim2019}.
\hfill$\square$
\end{pf}

\subsection{Ensuring Safety with Control Barrier Functions}\label{subsec:SOCP}

While Theorem~\ref{th:safety_existence} guarantees the existence of safe control inputs, we generally do not want to set $\bm{\nu}=\gamma\nabla^T_{\bm{x}}\psi(\bm{x})\bm{B}$ as used in the proof of this theorem. For example, when the task is to track a given reference trajectory $\bm{q}_d:\mathbb{R}\rightarrow\mathbb{R}^m$, we ideally want to use a tracking controller, e.g., 
\begin{equation}\label{eq:track_ctrl}
\bm{\nu}_{\mathrm{nom}}=\overset{(4)}{\dot{\bm{q}}_{d}}+\bm{L}\left(\bm{x}_{d}-\bm{x}\right),
\end{equation}
where $\bm{L} \in \mathbb{R}^{m \times n}$ is a stabilizing linear feedback gain  matrix and $\bm{x}_d=[\bm{x}_{d,1}\ \cdots\ \bm{x}_{d,4}]$ with
\begin{align}
    \bm{x}_{d,1}=\bm{q}_d,&& \bm{x}_{d,2}=\dot{\bm{x}}_{1}, &&\bm{x}_{d,3}=\dot{\bm{x}}_{2},&& \bm{x}_{d,4}=\dot{\bm{x}}_{d,3}.
\end{align}
A common approach to render existing control laws safe relies on the idea of modifying them towards safety in an optimization based fashion via 
\begin{subequations}\label{eq:qp}
\begin{align}
    \bm{\nu}^{*}(\bm{x})=&\arg\min\limits_{\bm{\nu}\in\mathbb{R}^m} \|\bm{\nu}_{\mathrm{nom}}-\bm{\nu}\|^2 \\
    &\text { s.t. \eqref{eqn:cbf} holds}.
\end{align}
\end{subequations}
When the dynamics of a system are known, such optimization problems can be formulated as quadratic programs, which can be efficiently solved. However, we cannot directly impose \eqref{eqn:cbf} as a constraint when using the learned dynamics since it depends on the unknown linearization errors $\bm{e}(\cdot)$ and $\bm{E}(\cdot)$. We circumvent this issue by exploiting the (probabilistic) linearization error bounds \eqref{eq:sigma_EGP}, \eqref{eq:e_GP}, \eqref{eq:sigma_Egamma}, \eqref{eq:e_gamma}.
This allows us to derive a (probabilistic) worst case of \eqref{eqn:cbf}, which can be efficiently included in a second-order cone program for ensuring the safety of arbitrary nominal control laws.
\begin{theorem}
Consider an elastic joint robot \eqref{eq:real_model} satisfying Ass.~\ref{as:boundednessproperty}. Assume that for its system description \eqref{eq:ctrl_aff_sys}, the functions $f_i(\cdot)$, $g_{ij}(\cdot)$ are sample functions from GPs, i.e., \eqref{eq:fprior}, \eqref{eq:gprior} hold. 
Then, the second-order cone program 
\begin{subequations}\label{eq:SOCP}
\begin{align}
    &\min\limits_{\bm{z}\in\mathbb{R}^{m+1}} \begin{bmatrix}-2\bm{\nu}_{\mathrm{nom}}^T& 1 \end{bmatrix}\bm{z}\\
    &\text{ s.t. } \|\bm{P}_i\bm{z}+\bm{q}_i\|\leq \bm{r}_i^T\bm{z}+s_i, \forall i=1,2
\end{align}
\end{subequations}
where 
\begin{align}
    \bm{P}_1&\!=\!\begin{bmatrix}2\bm{I}_m&\bm{0}\\ \bm{0}&1\end{bmatrix} & \bm{P}_2&\!=\!\begin{bmatrix}\gamma_E\|\nabla_{\bm{x}}^T\psi(\bm{x})\bm{B}\|\bm{I}_m&\bm{0}\\ \bm{0}&0\end{bmatrix}\\
    \bm{q}_1&\!=\!\begin{bmatrix}\bm{0}\\-1 \end{bmatrix} & \bm{q}_2\!&\!=\begin{bmatrix}\bm{0}\\0 \end{bmatrix}\nonumber\\
    \bm{r}_1&\!=\!\begin{bmatrix}\bm{0}\\1 \end{bmatrix}& \bm{r}_2&\!=\!\begin{bmatrix}\nabla_{\bm{x}}^T\psi(\bm{x})\bm{B}\\0 \end{bmatrix}\nonumber\\
    s_1&\!=\! 1& s_2&\!=\!\nabla_{\bm{x}}^T\!\psi(\bm{x})\bm{A}\bm{x}\!-\!\|\nabla_{\bm{x}}^T\!\psi(\bm{x})\bm{B}\|\bar{e}\!+\!\alpha(\psi(\bm{x}))\nonumber
\end{align}
and $\bm{z}=\begin{bmatrix}\bm{\nu}^T&t \end{bmatrix}^T$, is feasible for all $\bm{x}\in\mathbb{X}$ and its solution $\bm{\nu}^*$ ensures the safety of \eqref{eq:FelLi_pol} for elastic joint robots \eqref{eq:real_model} with probability of at least $1-(m+m^2)\delta$.
\end{theorem}

\begin{pf}
Due to Lemma~\ref{lem:cbf}, safety of the system is ensured if \eqref{eqn:cbf} holds. Since we do not know the functions $\bm{e}(\cdot)$, $\bm{E}(\cdot)$, we instead make use of the probabilistic worst case
\begin{align*}
    \bm{\xi}^T(\bm{A}\bm{x}+\bm{B}\bm{\nu})-\|\bm{\xi}^T\bm{B}\|(\gamma_E\|\bm{\nu}\|+\bar{e})\geq -\alpha(\psi(\bm{x})),
\end{align*}
where we use the shorthand notation $\bm{\xi}|\!=\!\nabla_{\bm{x}}\psi(\bm{x})$, define $\gamma_E(\bm{x})\!=\!\gamma_{E_{GP}}(\bm{x})$, $\bar{e}(\bm{x})\!=\!\bar{e}_{GP}(\bm{x})$ if \eqref{eq:GP_cond} holds and $\gamma_E(\bm{x})\!=\!\gamma_{E_\gamma}(\bm{x})$, $\bar{e}(\bm{x})\!=\!\bar{e}_{\gamma}(\bm{x})$ otherwise. As $\gamma_E(\bm{x})$ is a bound for $\|\bm{E}(\bm{x})\|$ with probability of at least $1\!-\!m^2\delta$ and $\bar{e}$ for $\|\bm{e}(\bm{x})\|$ with probability $1\!-\!(m^2\!+\!m)\delta$, \eqref{eq:rob_cbf_constr} implies the satisfaction of \eqref{eqn:cbf} with probability of at least $1\!-\!(m\!+\!m^2)\delta$. Therefore, given any nominal control input $\bm{\nu}_{\mathrm{nom}}$, safe control inputs $\bm{\nu}$ can be obtained by solving the optimization problem\looseness=-1
\begin{subequations}
\begin{align}\label{eq:rob_cbf_cost}
    &\min\limits_{\bm{\nu}\in\mathbb{R}^m} \|\bm{\nu}_{\mathrm{nom}}\!-\!\bm{\nu}\|^2  \\
    &\text { s.t. } \bm{\xi}^T\!(\bm{A}\bm{x}\!+\!\bm{B}\bm{\nu})\!-\!\|\bm{\xi}^T\!\bm{B}\|(\gamma_E\|\bm{\nu}\|\!+\!\bar{e})\!\geq\! -\!\alpha(\psi(\bm{x})).\!\!\!\label{eq:rob_cbf_constr}
\end{align}
\end{subequations}
It remains to reformulate this optimization problem into a second-order cone program. The cost function can be expressed in the required form by introducing a slack variable $t\in\mathbb{R}$, which yields the identity
\begin{subequations}
\begin{align}
    \min\limits_{\bm{\nu}\in\mathbb{R}^m} \|\bm{\nu}_{\mathrm{nom}}-\bm{\nu}\|^2=&\min\limits_{\bm{\nu}\in\mathbb{R}^m, t\in\mathbb{R}} -2\bm{\nu}^T\bm{\nu}_{\mathrm{nom}}+t\\
    &\text{ s.t. } \bm{\nu}^T\bm{\nu}\leq t.\label{eq:SOC1}
\end{align}
\end{subequations}
Constraint \eqref{eq:SOC1} can be formulated as the second order cone condition \citep{Alizadeh2003SecondorderCP}
\begin{align*}
    \left\|\begin{bmatrix}2\bm{I}_m&\bm{0}\\ \bm{0}&1\end{bmatrix}\begin{bmatrix}\bm{\nu}\\t \end{bmatrix}+\begin{bmatrix}\bm{0}\\-1 \end{bmatrix}\right\|\leq \begin{bmatrix}\bm{0}\\1 \end{bmatrix}\begin{bmatrix}\bm{\nu}\\t \end{bmatrix}+1,
\end{align*}
such that the cost \eqref{eq:rob_cbf_cost} has the form of a second order cone program. Finally, it can be straightforwardly seen that the constraint \eqref{eq:rob_cbf_constr} can be expressed as
\begin{align*}
    \big\| \gamma_E\|\bm{\xi}^T\bm{B}\|\bm{I}_m \bm{\nu} \big\|\leq \bm{\xi}^T\bm{B}\bm{\nu} + \bm{\xi}^T\bm{A}\bm{x}-\|\bm{\xi}^T\bm{B}\|\bar{e}+\alpha(\psi(\bm{x})),
\end{align*}
such that we can equivalently solve \eqref{eq:SOCP} for ensuring the safety of \eqref{eq:FelLi_pol} for elastic joint robots \eqref{eq:real_model} with probability of at least $1-(m+m^2)\delta$.\hfill$\square$
\end{pf}

\section{Numerical Evaluation}\label{sec:Eval}
\vspace{-0.1cm}

\setlength{\floatsep}{2pt}
\setlength{\textfloatsep}{9pt}

We evaluate the proposed approach on simulations of a two degree of freedom robot with elastic joints controlled with sampling rate of $100$\si{Hz}. We use unit lengths and masses for computing $\bm{M}(\cdot)$ and $\bm{C}(\cdot,\cdot)$ and set $\bm{J}=0.001\bm{I}_2$, $\bm{K}=\bm{I}_2$. Since $\bm{f}(\cdot)$ can be directly measured when no control inputs are applied, we consider a prior mean $\hat{\bm{f}}(\cdot)$ defined through a perturbation of the true robot parameters, while we assume $\hat{\bm{G}}(\bm{x})=\bm{0}$. We train a GP with squared exponential kernel using $786$ training samples on a uniform grid with observation noise variance $\sigma_{\mathrm{on}}=0.1$ and determine the hyperparameters using likelihood maximization \citep{Rasmussen2006}. In the GP error bounds \eqref{eq:GPboundf}, \eqref{eq:GPboundg}, we use $\beta=24$, which can be shown to ensure $\delta=0.05$ in Lemma~\ref{lem:GPbound} jointly for all times the GP is evaluated following the ideas of \citep{Lederer2019UniformEB}. The threshold for choosing the the GP-based controller in \eqref{eq:FelLi_pol} is set to $\zeta=0.95$ and 
we use $\bm{\pi}_{\gamma}(\cdot)$ with $\bar{\gamma}_{\bm{G}}=1640$, $\underline{\gamma}_{\bm{G}}=97$ for the back-up control law. As nominal control law we employ the linear tracking controller \eqref{eq:track_ctrl} with a manually tuned constant gain matrix
\begin{equation}
\bm{L} = \begin{bmatrix}
10^4 & 0 & 10^3  &  0 & 300 & 0 & 10 &0 \\ 
 0& 10^4 & 0 & 10^3 & 0 & 300 & 0  & 10
\end{bmatrix}
\end{equation}
and reference trajectories of the form $\bm{q}_{\mathrm{ref}}(t)=$\linebreak
$[\sin(\pi t/c)\!\!\ \cos(\pi t/c)]^T\!$, where the frequency is drawn from a uniform distribution $c\!\sim\!\mathcal{U}([4,100])$. The CBF is constructed using \eqref{eq:cbf_construct} with $b(\bm{q})\!=\!0.8\! - \!q_1\!$ and $\alpha(\bm{\psi})\!=\!16\psi$, such that the reference violates the safety condition.\looseness=-1

\begin{figure} 
    \vspace{0.1cm}
    \centering
    \definecolor{nicegreen}{RGB}{0,200,0} 
\definecolor{zblue}{RGB}{85, 129, 230}
\definecolor{zred}{RGB}{217,132,129}
\definecolor{zorange}{RGB}{248, 108, 60}
\definecolor{zpink}{RGB}{248, 92, 116}
\definecolor{zyellow}{RGB}{242, 202, 25}
\definecolor{zpurple}{RGB}{152, 60, 204}
\definecolor{zgreen}{RGB}{160, 204, 76}

	\pgfplotsset{width=10\columnwidth /10, compat = 1.13, 
		height = 43\columnwidth /100, grid= major, 
		legend cell align = left, ticklabel style = {font=\small},
		every axis label/.append style={font=\small},
		legend style = {font=\small\sffamily},title style={yshift=-7pt, font = \small} }

\def\file{files/switching_CBF_new.txt}
\def\filez{files/BUonly_CBF_new.txt}
\def\filey{files/GPonly_CBF_new.txt}
\centering
\begin{tikzpicture}
\node[name=plotvirtual] at (0,0) {\begin{tikzpicture}
\begin{axis}[
name=plot1,
height = 3.5cm, width =\columnwidth,
grid=none,
xmin=0, xmax=30,
ymin=-1.2, ymax=1.2,
ylabel={$q_1$},
 		legend columns=3,
 		legend style={at={(0.5,1.5)},anchor=north},
		xticklabels={,,}
]
        \addplot[black,thick,dashed]table[x=ts,y=cbf_bound]{\filez};
		\addplot[black,thick]table[x=ts,y=ref_1]{\filez};

		\addplot[cyan, very thick]table[x = ts_GP_c1,
					y = xs_GP_c1_1]{\file};
		\pgfplotsinvokeforeach{2,...,9}{
			\addplot[cyan, forget plot,very thick]table[x = ts_GP_c#1,
			y = xs_GP_c#1_1]{\file};
		}
		\addplot[orange, very thick]table[x = ts_BU_c1,
		y = xs_BU_c1_1]{\file};
		\pgfplotsinvokeforeach{2,...,8}{
			\addplot[orange, forget plot,very thick]table[x = ts_BU_c#1,
			y = xs_BU_c#1_1]{\file};
		}
		\addplot[red,thick,dashed]table[x=ts,y=xs_1]{\filez};
		\addplot[blue!70!black,thick,dashed]table[x=ts,y=xs_1]{\filey};
        
\end{axis}
\begin{axis}[
name=plot2,
at=(plot1.east), anchor=east, yshift=-2.0cm,
height = 3.5cm, width = \columnwidth,
grid=none,
xmin=0, xmax=30,
ymin=-1.2, ymax=1.2,
xlabel={$t$}, 
ylabel={$q_2$},
xlabel shift={-3pt},
]
\addplot[black,thick]table[x=ts,y=ref_2]{\filez};

\addplot[cyan, forget plot,thick]table[x = ts_GP_c1,
y = xs_GP_c1_2]{\file};
\pgfplotsinvokeforeach{2,...,9}{
	\addplot[cyan, forget plot,thick]table[x = ts_GP_c#1,
	y = xs_GP_c#1_2]{\file};
}
\addplot[orange, forget plot,thick]table[x = ts_BU_c1,
y = xs_BU_c1_2]{\file};
\pgfplotsinvokeforeach{2,...,8}{
	\addplot[orange, forget plot,thick]table[x = ts_BU_c#1,
	y = xs_BU_c#1_2]{\file};
}

\addplot[red,thick,dashed]table[x=ts,y=xs_2]{\filez};
\addplot[blue!70!black,thick,dashed]table[x=ts,y=xs_2]{\filey};

\end{axis}
\end{tikzpicture}};
\node[below=0.3cm of plotvirtual.south] (plot) {}; 

\node[above=5.25cm of plot.south, xshift=-3.15cm] (leftbottom) {};
\node[above=6.2cm of plot.south, xshift=3.93cm] (righttop) {};

\node[above=5.866cm of plot.south, xshift=-1.725cm] (constraint) {\small\sffamily \textcolor{black}{constraint}};
\node[above=5.866cm of plot.south, xshift=0.495cm] (ref) {\small\sffamily \textcolor{black}{reference}};
\node[above=5.787cm of plot.south, xshift=2.785cm] (GPonly) {\small\sffamily \textcolor{black}{$\bm{\pi}_{GP}(\cdot)$ only}};
\node[above=5.376cm of plot.south, xshift=-1.692cm] (backuponly) {\small\sffamily \textcolor{black}{$\bm{\pi}_{\gamma}(\cdot)$ only}};
\node[above=5.398cm of plot.south, xshift=1.6cm] (switch) {\small\sffamily \textcolor{black}{proposed switching control}};
\node[above=5.937cm of plot.south, xshift=-3.16cm] (constraint_line_left) {};
\node[above=5.937cm of plot.south, xshift=-2.315cm] (constraint_line_right) {};
\node[above=5.937cm of plot.south, xshift=-0.867cm] (ref_line_left) {};
\node[above=5.937cm of plot.south, xshift=-0.021cm] (ref_line_right) {};
\node[above=5.937cm of plot.south, xshift=1.21cm] (GPonly_line_left) {};
\node[above=5.937cm of plot.south, xshift=2.057cm] (GPonly_line_right) {};
\node[above=5.537cm of plot.south, xshift=-3.16cm] (backuponly_line_left) {};
\node[above=5.537cm of plot.south, xshift=-2.315cm] (backuponly_line_right) {};
\node[above=5.537cm of plot.south, xshift=-0.867cm] (switch_line_left) {};
\node[above=5.537cm of plot.south, xshift=-0.021cm] (switch_line_right) {};

\draw (leftbottom) rectangle (righttop);
\draw[black, thick, dashed] (constraint_line_left) -- (constraint_line_right);
\draw[black, thick] (ref_line_left) -- (ref_line_right);
\draw[blue!70!black, thick, dashed] (GPonly_line_left) -- (GPonly_line_right);
\draw[red, thick, dashed] (backuponly_line_left) -- (backuponly_line_right);
\draw[cyan, thick] (switch_line_left) -- (switch_line_right);
\draw[orange, thick] (switch_line_left)+(0.423cm,0cm) -- (switch_line_right);

\node[above=3.6cm of plot.south, xshift=-1.3cm] (GPactivetext) {\small\sffamily
\textcolor{orange}{$\bm{\pi}_{\gamma}(\cdot)$ active}};
\node[above=4.2cm of plot.south, xshift=2.8cm] (BUactivetext) {\small\sffamily
\textcolor{cyan}{$\bm{\pi}_{GP}(\cdot)$ active}};

\draw[Latex-, orange] (GPactivetext)+(0.2cm,0.95cm) -- (GPactivetext); 
\draw[Latex-, cyan] (BUactivetext)+(-1.9cm,-0.3cm) -- (BUactivetext.west); 
\end{tikzpicture}

    \vspace{-1.05cm}
    \caption{Comparison of the proposed approach for ensuring safety based on CBFs, which switches between GP-based (blue line) and a prior model based feedback linearization (orange line), with linearizing controllers purely based on the GP model (cyan dashed line) and prior model bounds (red dashed line).\looseness=-1
    }
    \label{fig:GP_CBF}
\end{figure}
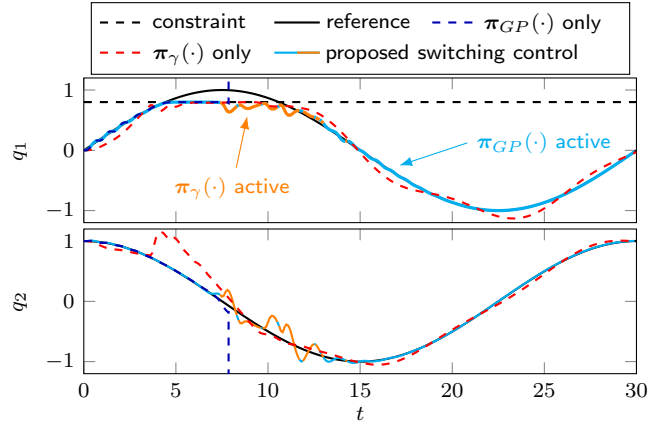

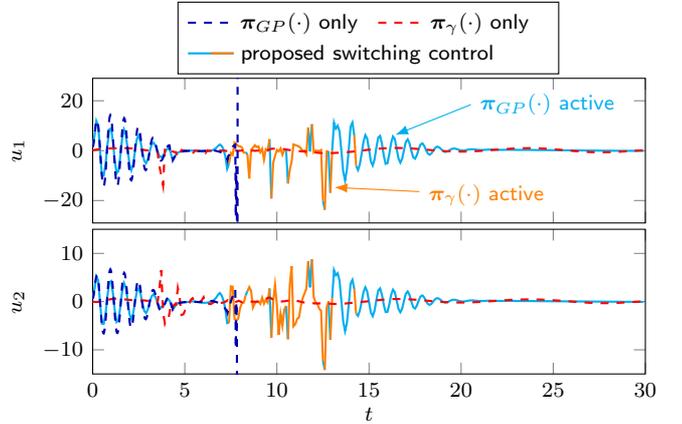
\begin{figure}[t]
    \vspace{0.1cm}
    \centering

\definecolor{nicegreen}{RGB}{0,200,0}

	\pgfplotsset{width=10\columnwidth /10, compat = 1.13, 
		height = 43\columnwidth /100, grid= major, 
		legend cell align = left, ticklabel style = {font=\small},
		every axis label/.append style={font=\small},
		legend style = {font=\small\sffamily},title style={yshift=-7pt, font = \small} }

\def\file{files/switching_CBF_new.txt}
\def\filez{files/BUonly_CBF_new.txt}
\def\filey{files/GPonly_CBF_new.txt}
\centering
\begin{tikzpicture}
\node[name=plotvirtual] at (0,0) {\begin{tikzpicture}
\begin{axis}[
name=plot1,
height = 3.5cm, width =\columnwidth,
grid=none,
xmin=0, xmax=30,
ymin=-29, ymax=29,
ylabel={$u_1$},
 		legend columns=2,
 		legend style={at={(0.5,1.53)},anchor=north},
		xticklabels={,,}
]

		\addplot[cyan, thick]table[x = ts_GP_c1,
					y = us_GP_c1_1]{\file};
		\pgfplotsinvokeforeach{2,...,9}{
			\addplot[cyan, forget plot, thick]table[x = ts_GP_c#1,
			y = us_GP_c#1_1]{\file};
		}
		\addplot[orange, thick]table[x = ts_BU_c1,
		y = us_BU_c1_1]{\file};
		\pgfplotsinvokeforeach{2,...,8}{
			\addplot[orange, forget plot, thick]table[x = ts_BU_c#1,
			y = us_BU_c#1_1]{\file};
		}
		\addplot[red,thick, dashed]table[x=ts,y=us_1]{\filez};
		\addplot[blue!70!black,thick, dashed]table[x=ts,y=us_1]{\filey};



\end{axis}
\begin{axis}[
name=plot2,
at=(plot1.east), anchor=east, yshift=-2.0cm,
height = 3.5cm, width = \columnwidth,
grid=none,
xmin=0, xmax=30,
ymin=-15, ymax=15,
xlabel={$t$}, 
ylabel={$u_2$},
xlabel shift={-3pt},
]

\addplot[cyan, forget plot,thick]table[x = ts_GP_c1,
y = us_GP_c1_2]{\file};
\pgfplotsinvokeforeach{2,...,9}{
	\addplot[cyan, forget plot, thick]table[x = ts_GP_c#1,
	y = us_GP_c#1_2]{\file};
}
\addplot[orange, forget plot,thick]table[x = ts_BU_c1,
y = us_BU_c1_2]{\file};
\pgfplotsinvokeforeach{2,...,8}{
	\addplot[orange, forget plot,thick]table[x = ts_BU_c#1,
	y = us_BU_c#1_2]{\file};
}

\addplot[red,thick,dashed]table[x=ts,y=us_2]{\filez};
\addplot[blue!70!black,thick,dashed]table[x=ts,y=us_2]{\filey};



\end{axis}
\end{tikzpicture}};
\node[below=0.29cm of plotvirtual.south] (plot) {}; 

\node[above=5.25cm of plot.south, xshift=-2.057cm] (leftbottom) {};
\node[above=6.2cm of plot.south, xshift=2.95cm] (righttop) {};

\node[above=5.778cm of plot.south, xshift=-0.465cm] (GPonly) {\small\sffamily \textcolor{black}{$\bm{\pi}_{GP}(\cdot)$ only}};
\node[above=5.772cm of plot.south, xshift=1.913cm] (backuponly) {\small\sffamily \textcolor{black}{$\bm{\pi}_{\gamma}(\cdot)$ only}};
\node[above=5.387cm of plot.south, xshift=0.44cm] (switch) {\small\sffamily \textcolor{black}{proposed switching control}};
\node[above=5.927cm of plot.south, xshift=0.453cm] (backuponly_line_left) {};
\node[above=5.927cm of plot.south, xshift=1.298cm] (backuponly_line_right) {};
\node[above=5.521cm of plot.south, xshift=-2.05cm] (switch_line_left) {};
\node[above=5.521cm of plot.south, xshift=-1.201cm] (switch_line_right) {};
\node[above=5.927cm of plot.south, xshift=-2.05cm] (GPonly_line_left) {};
\node[above=5.927cm of plot.south, xshift=-1.201cm] (GPonly_line_right) {};

\draw[] (leftbottom) rectangle (righttop);
\draw[blue!70!black, thick, dashed] (GPonly_line_left) -- (GPonly_line_right);
\draw[red, thick, dashed] (backuponly_line_left) -- (backuponly_line_right);
\draw[cyan, thick] (switch_line_left) -- (switch_line_right);
\draw[orange, thick] (switch_line_left)+(0.423cm,0cm) -- (switch_line_right);

\node[above=3.5cm of plot.south, xshift=2.0cm] (GPactivetext) {\small\sffamily
\textcolor{orange}{$\bm{\pi}_{\gamma}(\cdot)$ active}};
\node[above=4.7cm of plot.south, xshift=2.8cm] (BUactivetext) {\small\sffamily
\textcolor{cyan}{$\bm{\pi}_{GP}(\cdot)$ active}};

\draw[Latex-, orange] (GPactivetext)+(-2.0cm,0.1cm) -- (GPactivetext); 
\draw[Latex-, cyan] (BUactivetext)+(-2.0cm,-0.4cm) -- (BUactivetext.west);

\end{tikzpicture}
    \vspace{-1.4cm}
    \caption{Comparison of the control inputs $u_i$ resulting from the proposed feedback linearization, which switches between a GP model (blue line) and a prior model (orange line), with controllers purely based on the GP model (cyan dashed line) and prior model bounds (red dashed line).}
    \label{fig:GP_CBF_controls}
\end{figure}

The resulting trajectories of the proposed approach in comparison to using only a GP model or only the prior model bound $\bar{\gamma}_G$ in feedback linearization are exemplarily illustrated for a reference with $c=15$ in Fig.~\ref{fig:GP_CBF}. When using only the GP model, the robustified CBF condition becomes infeasible, leading to unpredictable behavior and divergence after~$8$s. The feedback linearization $\bm{\pi}_{\gamma}(\cdot)$ alone is safe, but yields poor tracking accuracy. In contrast, the proposed switching approach ensures a high accuracy using the GP whenever possible, but activates $\bm{\pi}_{\gamma}(\cdot)$ when necessary to preserve the feasibility of the SOCP. The reason for this behavior becomes clear when looking at the corresponding control input signals, which are illustrated in Fig.~\ref{fig:GP_CBF_controls}. When approaching $\approx 8$s, the GP-based feedback linearization yields an infeasible CBF condition, the control input extremely grows. This effectively causes the pure GP-based controller to fail. While the back-up controller $\bm{\pi}_{\gamma}(\cdot)$ based on the prior model bound is safe, it results in comparatively small control amplitudes. Thereby, it is not capable of achieving a high tracking accuracy. In contrast, the proposed switching between the control laws results in inputs similar to the GP-based controller but avoids excessive magnitudes due to infeasibility. 

As depicted in Tab.~\ref{tb:margins}, these advantages are not just limited to the particular example reference, but also hold for randomly sampled parameters $c$. The proposed approach is capable of significantly reducing the average tracking error, while at the same time avoiding any infeasible optimization problems \eqref{eq:SOCP}. This clearly demonstrates the high performance and safety achieved through the combination of prior model bounds and a learned GP model.\looseness=-1


\vspace{-0.05cm}
\section{Conclusion}\label{sec:conc}
\vspace{-0.1cm}

In this paper, we have proposed a novel approach for ensuring the safe control of elastic joint robots by combining GP regression with control barrier functions. We learn a model of the robot dynamics with GP regression, which is employed in a feedback linearizing controller. To ensure the feasibility of CBF conditions, we switch to a feedback linearization based on prior model bounds whenever necessary. We reformulate the CBF conditions as second-order cone constraints so that they can be efficiently enforced using on-line optimization. The effectiveness of the approach is demonstrated in simulations.

\begin{table}[!t]
	\caption{Comparison of the proposed switching strategy with approaches based on either prior model bounds or a GP model for $100$ random reference trajectories. Due to infeasibilities for the purely GP-based control law, no mean squared error can be provided.}\label{tb:margins}
	\vspace{0.1cm}
	\centering
	\begin{footnotesize}
	\setlength{\tabcolsep}{3.75pt}
	\vspace{-0.35cm}
			\begin{tabular}{l c c c}
				\toprule
				  &  proposed approach & prior bounds & GP only\\
				\midrule
				mean squared error & $\bm{0.0370}$ & $0.0754$ & --- \\
				\# infeasibilities & $\bm{0}$ & $\bm{0}$ & $77$\\
				\bottomrule
			\end{tabular}
		
	\end{footnotesize}
	\vspace{0.25cm}
\end{table}

\vspace{-0.05cm}
\bibliography{ifacconf}    

\end{document}